\documentclass[pra, twosides,preprintnumbers, twocolumn]{revtex4}
\usepackage{amssymb}

\usepackage{graphicx, amsmath, amssymb}

\begin{document}

\preprint{SB/F/06-355}

\title{Concurrence and negativity as distances}
\author{D.F. Mundarain and J. Stephany}
\affiliation{ \
Departamento de F\'{\i}sica, Secci\'{o}n de Fen\'{o}menos \'{O}pticos, Universidad Sim\'{o}n
Bol\'{\i}var,
Apartado Postal 89000, Caracas 1080A, Venezuela }

\begin{abstract}
In this report  we consider the three dimensional subset of the space of states of two qubits that may be written in the so called standard form. For those states we  show that different measures of entanglement, specifically concurrence, negativity and the
 Hilbert-Schmidt distance  are proportional to  the euclidean distance between the point representing the state in the three dimensional parameter space  and the set of separable states.
\end{abstract}

\maketitle

\section{Introduction}
The possibility of preparing quantum systems of few particles in entangled states may be the more characteristic aspect of quantum dynamics. It is the one which is farthest of our classical intuition  and the one which tests with more exigence our comprehension of the quantum world \cite{EinAPR1935,HorRHH2007}. Recently entanglement and its counterpart decoherence gain also importance as a key element in the discussion about definitions and applications of quantum computing, quantum cryptography and quantum teleportation\cite{BenCBC1993}.

The quantification of the degree of entanglement\cite{HorRHH2007,PleMV2005} corresponding to a  quantum state is of great importance both for the understanding of fundamental aspects of quantum mechanics and for possible applications. For systems with many degrees of freedom  developments had stressed approximate methods for the separation of the entangled component of a given configuration. \cite{LewMS1998,PleMV2005,HorRHH2007} and a complete characterization has not been achieved. But for two qubits systems the works of   Peres, Horodecki, Hill, and  Wootters \cite{PerA1996,HorMHH1996,HilSW1997,WooW1998} have established the base for a complete discussion of entanglement in terms of  algebraic properties of the density matrix.  The so-called Peres-Horodecki criterium \cite{PerA1996,HorMHH1996} on one hand and  concurrence as defined by Wootters  \cite{WooW1998} in the other give the tools  to distinguish completely entangled states from separate states. Other entanglement measures e.g entanglement of formation \cite{BenCDS1996} and the Hilbert Schmidt distance \cite{WitCT1999,OzaM2000,BertRNT2002} also work consistently for these systems.  More recently  in Refs.\cite{VerFDD2001,LeiJMO2006} some interesting geometrical aspects  of the  separable and entangled sectors of the space of physical states were discussed.

In this paper we discuss further  two qubits systems. For states in the three dimensional subset of the physical space that may be written in the standard  form \cite{LeiJMO2006} (see Eq. (\ref{standard}) below) we show the specific geometrical relation between  concurrence  \cite{WooW1998}, negativity \cite{ZycKHS1998,VerFAD2001}, and Hilbert Schmidt distance  \cite{WitCT1999,OzaM2000,BertRNT2002}  and the euclidean distance between the point representing the state in the three dimensional parameter space  and the set of separable states.

\section{Entanglement measures}
Let us first discuss very briefly the definitions and properties of the entanglement measures that we need for our discussion below.
For a pure state of a composed system with subsystems $A$ and $B$, entanglement is defined as the entropy of either of the two subsystems . The entanglement of formation of a mixed state $\rho$ defined as  the average entanglement of the pure states in its decomposition minimized over all decompositions \cite{BenCDS1996} is a good measure of the entanglement of the system. It can represented as a monotonically increasing function of concurrence introduced in \cite{HilSW1997,WooW1998} which then can also be taken as an entanglement measure. For two qubits, given a  density matrix $\rho $ , concurrence \cite{WooW1998} is calculated in terms of the eigenvalues $R_1, R_2, R_3,R_4$  of the
matrix  $R$ defined by,
\begin{equation}
R =  \rho \,\, \sigma_y \, \otimes\, \sigma_y \,\,\rho^{*}\,\,\sigma_y\,\otimes\, \sigma_y
\end{equation}
It is given by
\begin{equation}
C = {\it max} \{0, 2 \sqrt{R_m}-(\sqrt{R_1}+\sqrt{R_2}+\sqrt{R_3}+\sqrt{R_4})\}
\end{equation}
where
\begin{equation}
R_m = {\it max} \{R_1, R_2, R_3,R_4\}
\end{equation}
For maximally entangled states concurrence is 1 wether for separable states is zero.

The Peres-Horodeky \cite{PerA1996} condition  establishes that a matrix is separable if its partial transposed matrix has only positive eigenvalues, ie. if the partial transposed matrix belongs to the physical space of the two particles. For $2\times 2$ systems this condition is also sufficient to characterize a separable state \cite{HorMHH1996}. In fact it happens that for these systems the partial transpose of a nonseparable state has one negative eigenvalue $\lambda_N$. Negativity \cite{ZycKHS1998,VerFAD2001} may then be defined (for two qubits systems)  as twice the absolute value of this negative eigenvalue.
\begin{equation}
E_N= 2 max(0,-\lambda_N)
\end{equation}
and  may be used also to quantify entanglement. Finally another entanglement measure we will use in what follows is the Hilbert Schmidt \cite{WitCT1999,OzaM2000,BertRNT2002} distance given by
\begin{equation}
\label{HS}
D_{HS}(\rho) = \min_{\omega \,\in S} \|\, \omega-\rho \, \|_2 \,
\end{equation}
where $S$  is the set of separable states and
\begin{equation}
\|\, \omega- \rho \, \|_2^2 \, =\, \mbox{Tr }\big( \,\omega^2 + \rho^2 - 2 \, \sqrt{\rho} \;
\rho \, \sqrt{\rho} \, \big) \ .
\end{equation}

\section{The geometry of the two qubits Hilbert space}
\label{sec3}
For two qubits systems which are the ones we are concerned in this note,
density matrices are represented in terms of the Pauli matrices ($\sigma_\mu\mapsto\sigma_0=\mathbb{I},\sigma_i$) in the form,
 \begin{equation}
\label{general}
\rho = \frac{1}{4} \left( \sum_{i=0}^3\,  r_{\mu\nu} \,\sigma_\mu^A \otimes \sigma_\nu^B \right)
\end{equation}
Local unitary transformations  do not modify the degree of entanglement of a given state. On the other hand entanglement may be modified by filtering operations. Verstraete {\it et al} \cite{VerFDD2001} show that filtering operations on two qubits correspond to Lorentz transformations on the real parametrization (\ref{general}) of the density matrix. For filtering operations of the form,
\begin{equation}
\rho'=\frac{(A\otimes B)\rho(A\otimes B)^\dagger}{Tr[(A\otimes B)\rho(A\otimes B)^\dagger}]
\end{equation}
they show that concurrence transforms as
 \begin{equation}
 \label{cprime}
C'=C\frac{|det(A)||det(B)|}{Tr[A^\dagger A\otimes B[ \dagger B \rho}]
\end{equation}
Since it appears a normalization factor in this relation, it links the changes in entanglement to the non-locality of the transformation. Leinaas {\it et al} \cite{LeiJMO2006} working  on this characterization of the filtering operations  then showed that by means of
such transformations any density matrix of the form (\ref{general}) can be transformed to the standard form
\begin{equation}
\label{standard}
\rho = \frac{1}{4} \left( 1+ \sum_{i=1}^3\,  r_i \,\sigma_i^A \otimes \sigma_i^B \right)
\end{equation}
with separable states mapping on separable states. These results suggests that although restricted the space of states written in the standard form should be nevertheless useful to understand partially the properties of the whole set of states particulary in relation with separability.

Following \cite{VerFDD2001,LeiJMO2006} we consider the geometrical space $V$ with coordinates $r_i$.
The eigenvalues of the density matrix are given by
\begin{equation}
\rho_1 = \frac{1}{4}\left( 1+r_x-r_y+r_z \right)
\end{equation}
\begin{equation}
\rho_ 2=\frac{1}{4}\left( 1-r_x+r_y+r_z \right)
\end{equation}
\begin{equation}
\rho_3 =\frac{1}{4}\left( 1+r_x+r_y-r_z \right)
\end{equation}
\begin{equation}
\rho_4=\frac{1}{4}\left( 1-r_x-r_y-r_z \right)
\end{equation}
and are necessarily  positive. The equations $\rho_i=0$  define planes in $V$ below each one of  the matrices defined by (\ref{standard}) no longer is an acceptable density matrix  for the two particles. Taken together, these  four planes define a tetrahedron with vertices at $(r_x,r_y,r_z) =
\left\{ (-1,1,1), (1,-1,1),(1,1,-1), (-1,-1,-1) \right\}$ \cite{LeiJMO2006}. The  points inside
the tetrahedron correspond to physical states of the two particles and the points
outside do not belong to the physical space. This tetrahedron may be called the physical tetrahedron. In the vertices of the physical tetrahedron are located the points representing the states of maximal entanglement or Bell states.

 The  partial transposition used to implement  Peres-Horodeky criterium produces the following changes on a density  matrix  written in the standard form,
\begin{eqnarray}
 r_x &\rightarrow &r_x \nonumber \\
r_y &\rightarrow &-r_y \nonumber \\
 r_z &\rightarrow &r_x
\end{eqnarray}
This means that under partial transposition the physical tetrahedron is transformed
to another tetrahedron, the Peres-Horodeky tetrahedron which has vertices at $(r_x,r_y,r_z) =
\left\{ (-1,-1,1), (1,1,1),(1,-1,-1), (-1,1,-1) \right\}$.

The intersection between these two tetrahedrons which is an octahedron is the set of separable states which may be written in the standard form.

\section{Concurrence and negativity  as distances}
\label{sec4}
The four smaller tetrahedrons, one near  each of the vertices of the physical
tetrahedron make together the set of entangled states that may be written in the standard form.  For a state in this set, the  natural
way to quantify entanglement  is taking the euclidian distance from the state to the octahedron
of separable matrices. In order to measure this distance one must determine first in
which corner  of the physical tetrahedron is the state located and then simply
calculate the distance to the octahedron by calculating the distance to the
respective plane of the Peres-Horodeky tetrahedron. For example, the Euclidean
distance of the point $(r_x,r_y,r_z) = (1,-1,1),$ which corresponds to a maximal entangled state, to the octahedron is equal to the distance  from this point to the
plane $x-y+z-1=0$ which is the nearest side of the Peres-Horodeky tetrahedron
to the point under study. In this case one gets $D(1,-1,1)= 2/\sqrt{3}$. For a
general point near the same vertex the distance to the octahedron is computed using elementary vector calculus to be
\begin{equation}
\label{eq2}
D(r_x,r_y,r_z) = \frac{1}{\sqrt{3}} \left( r_x-r_y+r_z-1\right)
\end{equation}
To compare with concurrence we observe that for matrices written in the standard form one has
\begin{equation}
R =  \rho^2
\end{equation}
then the eigenvalues of R are $\{\rho_1^2,\rho_2^2,\rho_3^2,\rho_4^2\}$.
The concurrence is given by,
\begin{equation}
C = {\it max} \{0, 2 \rho_m-(\rho_1+\rho_2+\rho_3+\rho_4)\}
\end{equation}
where
\begin{equation}
\rho_m  = {\it max} \{\rho_1,\rho_2,\rho_3,\rho_4\}
\end{equation}
For the point  $(r_x,r_y,r_z) = (1,-1,1),$ of our example one has,
\begin{equation}
 \rho_1= 1,\rho_2= \rho_3= \rho_4= 0
\end{equation}
Then $\rho_m= \rho_1 =1$ and  concurrence is 1 as should be  for a maximal entangled state. For points in the small tetrahedron near this vertex  $\rho_m= \rho_1 \neq 1$ and  concurrence takes the value,
\begin{equation}
C =  2 \rho_1-(\rho_1+\rho_2+\rho_3+\rho_4)\}.
\end{equation}
Subtituting the values of the eigenvalues one obtains:
\begin{equation}
\label{eq1}
C =  \frac{1}{2}\left( 1+r_x-r_y+r_z \right) -1 =\frac{1}{2} \left( r_x-r_y+r_z-1 \right)
\end{equation}
Then from Eq. (\ref{eq2}) and (\ref{eq1}) one obtains the following relation between the concurrence and the euclidean distance,
\begin{equation}
 D = \frac{2 C}{\sqrt{3}}
\end{equation}

To obtain the relation between (\ref{eq2}) and negativity we note that
the eigenvalues  of the partial transposed matrix are given by \cite{LeiJMO2006},
\begin{equation}
\rho_1^{PT}= \frac{1}{4}\left( 1+r_x+r_y+r_z \right)
\end{equation}
\begin{equation}
\rho_ 2^{PT}=\frac{1}{4}\left( 1-r_x-r_y+r_z \right)
\end{equation}
\begin{equation}
\rho_3^{PT}=\frac{1}{4}\left( 1+r_x-r_y-r_z \right)
\end{equation}
\begin{equation}
\label{neg}
\rho_4^{PT}=\frac{1}{4}\left( 1-r_x+r_y-r_z \right)
\end{equation}
Taking again  the point $(1,-1,1)$ one gets $\rho_1^{PT}=\rho_2^{PT}=\rho_3^{PT}=1/2$  and $\rho_4^{PT}=-1/2$.  For points in the corner near this vertex the negative eigenvalue of the partial transposed matrix will still be $\rho_4^{PT}$. From (\ref{neg}) one obtains the equivalence between concurrence and negativity for states written in the standard form,
\begin{equation}
 C = -2 \, \, \rho_4^{PT}=E_N
\end{equation}
Hence negativity is also proportional to the euclidean distance (\ref{eq2}). For points in the other three tetrahedrons of entangled states analogous  results are proven in the same way. Finally using the same tools it is straightforward to check that the Hilbert Schmidt distance (\ref{HS}) also is proportional  the euclidean distance (\ref{eq2}).

\section{Conclusion}
\label{sec5}
In this report  we consider two qubits systems as characterized by Leinaas {\it et al} in Ref.\cite{LeiJMO2006}. We show that for elements of the three dimensional set of states which can be represented in  standard form \cite{LeiJMO2006}  concurrence  \cite{WooW1998} is given by $\sqrt{3}/2$ times the geometrical distance from the point representing the state in the parameter space  to the set of separable states in the same representation.  For states in this set negativity as defined in Eq.(\ref{neg}) and concurrence take the same value and also proportional to the Hilbert-Schmidt distance. As an application of this computation we note that in the space of states written in the standard we are allowed to choice sets of states with the same entanglement by direct geometrical inspection. It should be interesting to combine the properties of the filtering operation (see Eq.(\ref{cprime})) and the geometrical aspects discussed in this note to characterize the entanglement of general states which are not written in the standard form.

\section{Acknowledgements}
This work was supported by Did-Usb Grant Gid-30 and by Fonacit Grant No G-2001000712.

\bigskip

\end{document}